\documentclass[10pt,twocolumn,letterpaper]{article}

\usepackage{wacv}              
\definecolor{wacvblue}{rgb}{0.21,0.49,0.74}
\usepackage[pagebackref,breaklinks,colorlinks,allcolors=wacvblue]{hyperref}
\usepackage{multirow}
\def\wacvPaperID{3141} 
\def\confName{WACV}
\def\confYear{2026}

\title{Overcoming Data Scarcity and Confidentiality in Hardware Assurance via Synthetic Generation}

\author{Gijung Lee, Ronald Wilson, Damon L. Woodard, Domenic Forte\\
Florida Institute of National Security, University of Florida, Gainesville, Florida, USA\\
{\tt\small \{lee.gijung, ronaldwilson\}@ufl.edu \{dwoodard, dforte\}@ece.ufl.edu}}

\begin{document}

\maketitle






\begin{abstract}
Hardware assurance relies on scanning electron microscopy (SEM) to verify nanoscale structures, but assembling the large, high-quality datasets required for automated analysis is impeded by time-intensive acquisition and strict intellectual property (IP) constraints on proprietary designs. We propose a privacy-preserving pipeline that secures IP by heavily distorting the functional design while generating a visually realistic synthetic dataset from a small set of initial examples. A StyleGAN first learns the distribution of hardware layout masks to generate novel, macroscopically varied structures. Subsequently, a conditional GAN (Pix2PixHD) translates these masks into realistic SEM images that preserve authentic textures and noise. The primary finding of this work is that a segmentation model trained exclusively on this synthetic data not only demonstrates a successful ``sim-to-real'' transfer to real images but also outperforms a baseline model trained on the limited real dataset. Because the underlying synthetic layouts are demonstrably novel and reproduce none of the specific proprietary routing of the original design, deploying the final segmentation model mitigates the risk of exposing sensitive IP to attacks like gradient inversion and membership inference, providing a highly secure, high-performance solution for hardware assurance.
\end{abstract}
\section{Introduction}
\label{sec:introduction}

Semiconductors are ubiquitous. They are found in almost everything ranging from everyday conveniences to critical infrastructure and mission critical systems, incentivizing bad actors to compromise their integrity. Ensuring trust in semiconductors requires a complete teardown of the integrated circuit (IC), recovery of underlying patterns in the die and verification of these patterns against design layouts with verifiable provenance in a process called hardware assurance (HA). AI plays a critical role in automatically extracting these patterns via computer vision techniques like segmentation. The key issue that sets apart this segmentation process is the nature of the data involved. Integrated Circuits (ICs) contain sensitive intellectual property (IP) representing significant research and development investment. Accidental disclosure of this design data could have widespread economic and security impacts. For a manufacturer, it could enable competitors to reverse engineer and replicate proprietary designs, eroding their market advantage. For users, it could lead to a flood of unauthorized, counterfeit chips that may be unreliable or fail in critical applications, posing significant safety and security risks. The sensitive nature of the data coupled with wide variations in semiconductor design patterns due to changes in technology node and manufacturer or changes in image acquisition settings in the modality---Scanning Electron Microscopy (SEM) in this case---can cause significant generalization issues in segmentation algorithms. Further, the lack of availability of real large-scale SEM image datasets for semiconductors also prevents the adoption of latest advancements in the community. Synthetic image generation presents a viable alternative but existing approaches are plagued with issues like lack of sufficient realism due to misalignment with imaging settings \cite{wilson2021refics}, generation capability limited to a select few layers of the device (like metal layers) and requirement for access to often unavailable ground-truth design layout data \cite{tee2023integrated}. 

To remedy these challenges, we propose a novel framework that utilizes a few seed images from the assurance process to generate synthetic images calibrated to the imaging modality in a privacy-preserving workflow, such that the sharing of generated data and trained models will no longer pose a security challenge. The core contributions and benefits of our proposed workflow are summarized as follows:
\begin{itemize}
    \item \textbf{A Generative Pipeline for Data Scarcity:} We introduce a multi-stage generative pipeline utilizing StyleGAN and Pix2PixHD to create high-fidelity, varied synthetic SEM datasets from a minimal number of initial real examples, effectively overcoming the bottlenecks of manual annotation and data scarcity.
    
    \item \textbf{Privacy Preservation via Design Distortion for Real-World Deployment:} In a practical, real-world workflow, a data owner (such as a foundry or secure lab) can generate this synthetic dataset and train the segmentation model entirely in-house. The trained segmentation model can then be freely shared with untrusted third-party testing facilities, external researchers, or offshore teams for automated hardware assurance. We define this pipeline as privacy-preserving because it heavily distorts the proprietary electrical design while maintaining necessary visual features. Even if the shared public model is subjected to advanced extraction attacks, the recovered layouts reproduce none of the specific proprietary routing of the original design and would, at most, correspond to a different layout rather than the protected one, making them meaningless for reverse engineering the original IP.
    
    \item \textbf{Comprehensive Layer Applicability:} Advancing beyond previous synthetic approaches, our method successfully models and synthesizes complex semiconductor structures across multiple physical levels, including metal, polysilicon, and doped layers.
    
    \item \textbf{Successful Sim-to-Real Transfer:} We demonstrate that a deep learning segmentation model (U-Net) trained exclusively on our synthetic data achieves highly accurate segmentation on real SEM images, even outperforming a baseline model trained on the limited real dataset.
\end{itemize}
\section{Background}
\label{sec:background}
\subsection{Generative Adversarial Networks (GANs)}
GANs~\cite{goodfellow2020generative} learn to mimic the natural image distribution by training a generator to produce samples that a discriminator cannot distinguish from real images. This adversarial setup has driven advances in high‑quality image synthesis \cite{arjovsky2017wasserstein, radford2015unsupervised, zhao2016energy}, representation learning \cite{salimans2016improved}, and controllable image manipulation \cite{zhu2016generative}. GAN‑generated data also boost object‑detection training \cite{li2017perceptual} and have been extended to video generation \cite{mathieu2015deep, tulyakov2018mocogan, vondrick2016generating}. Recent work pushes resolution and realism even further \cite{wang2018high}.

\vspace{0.2em}
\noindent\textbf{Style Based Generator:} A style-based generator, introduced in the StyleGAN framework \cite{karras2019style}, is a generative architecture that separates high-level attributes (“style”) from low-level details during the synthesis process. Unlike traditional generators that map latent codes directly to images, style-based generators use an intermediate latent space and adaptive instance normalization (AdaIN) to control image features at different levels of the network. 
This design enables fine-grained manipulation of outputs, where global structure and local details can be adjusted independently. As a result, style-based generators produce high-fidelity and interpretable results, and they have become a standard approach for image generation and reconstruction tasks where both realism and controllability are important.

\vspace{0.2em}
\noindent\textbf{Image-to-Image Translation}: Adversarial learning has become a cornerstone of image-to-image translation, a task that learns a mapping between paired images from a source and target domain \cite{isola2017image}. The success of this approach is due to the adversarial loss \cite{goodfellow2020generative}, which avoids the blurry results common with L1 loss \cite{isola2017image, johnson2016perceptual} by using a discriminator as a learned loss function that adapts to the data. A key implementation of this is the pix2pix framework \cite{isola2017image}, which leverages conditional GANs \cite{mirza2014conditional} for diverse applications like translating maps to satellite views. This was later extended by Pix2PixHD \cite{wang2018high} for high-resolution tasks, which employs a coarse-to-fine generator and multi-scale discriminators to synthesize photorealistic images for applications like semantic-to-image translation.

\subsection{Segmentation in Hardware Assurance (HA)}
IC reverse engineering (RE) plays a crucial role in hardware assurance by reconstructing circuit layouts from images of physical structures. This process is essential for verifying designs, detecting unauthorized modifications, and ensuring the integrity of microelectronics. This process relies heavily on image segmentation, particularly when working with SEM images to separate key circuit components. Image segmentation is the process of partitioning an image into meaningful areas of interest, facilitating easier analysis and interpretation of its contents by identifying and isolating objects or areas of interest. In the case of hardware RE, segments would include doped regions, polysilicon or metal gates, contacts and vias, and metal interconnects. Accurate segmentation remains a challenge due to variations in image quality, noise, and complex semiconductor structures. While various segmentation techniques exist, deep learning (DL)-based approaches such as U-Net have shown superior performance in extracting complex semiconductor structures compared to traditional image processing methods \cite{wilson2021refics, kalber2021u, qiao2025ra}.

\subsection{Synthetic Data Generation in HA}
In hardware assurance (HA), accurate segmentation of IC images is essential for detecting structural and functional errors. However, creating large annotated datasets is expensive, time-consuming, and raises privacy concerns, since real IC layouts may contain proprietary or security-sensitive information. Synthetic datasets provide a promising solution by enabling model training without exposing sensitive design data.
IC Mask–GAN \cite{tee2023integrated} generates synthetic masks with controllable parameters such as line width and roughness, but the masks are restricted to the metal layer. These masks are then translated into realistic IC images, and segmentation models are trained with a mix of the real dataset and the generated data. Conversely, Shape-Consistent Image Translation \cite{tee2025integrated} bypasses the need for real images to train a segmentation model, relying instead on real masks to generate synthetic IC images that preserve geometric fidelity.
Our method advances beyond these approaches. It is not restricted to the metal layer but can be applied to doped, metal, and polysilicon layers. In addition, it does not require real datasets to train a segmentation model, and once training is complete, it does not require real masks or real datasets to generate synthetic IC images. This makes our method both more flexible across IC layers and inherently privacy-preserving, since it minimizes reliance on sensitive real-world layouts throughout the entire workflow.

\section{Methodology}
\label{sec:methodology}
\begin{figure*}[t]
\centering
\includegraphics[width=0.9\linewidth, height=0.45\linewidth]{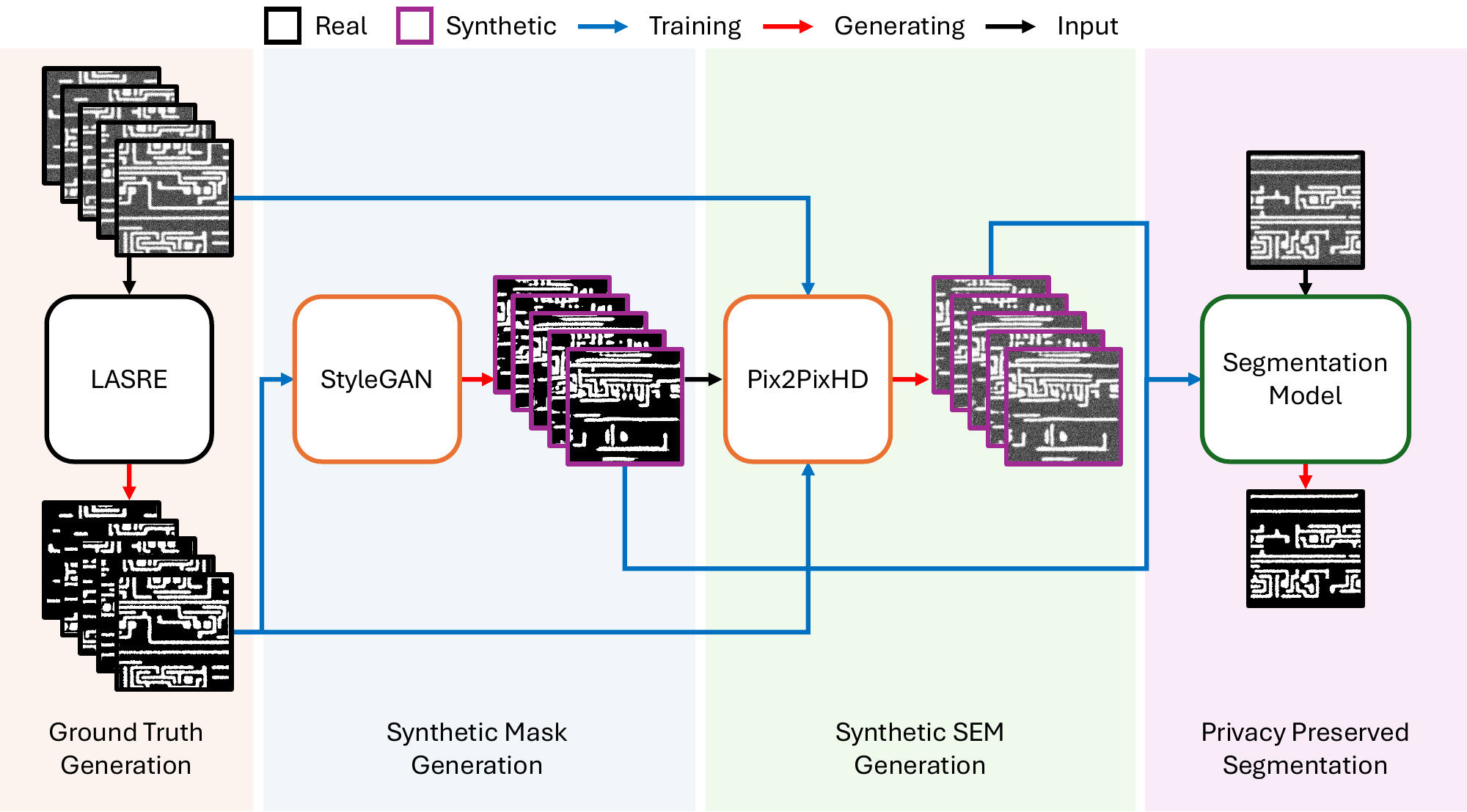}
\caption{Privacy Preserved pipeline for hardware assurance.\newline
GAN models are trained with only real datasets and a segmentation model is trained with only synthetic dataset. \newline
Only the trained segmentation model is shared with untrusted third parties.}
\label{fig:process}
\end{figure*}

To address the challenges of data scarcity and privacy, we developed the three-phase methodology detailed in this section. Figure \ref{fig:process} describes the overall process of our pipeline. We will first describe the phases of synthetic data generation and the segmentation model training, and conclude by detailing the quantitative methods used to validate the privacy-preserving capabilities of our approach.

\subsection{Dataset and Environment Setup}

A smart card IC was deprocessed to generate a dataset of real SEM images. The images were acquired using a secondary-electron (SE) detector at an accelerating voltage of 5\,keV, a field of view of 200\,$\mu$m, and a dwell time of 10\,$\mu$s per pixel. First, a large composite image (1076 × 4096 pixels) was produced. From this source, 505 smaller images (256 × 256 pixels) were extracted to form the final dataset. To generate the corresponding ground truth, we applied the LASRE unsupervised segmentation algorithm \cite{wilson2020lasre}. This approach is advantageous as it creates segmentation masks from the source images themselves, meaning our application only required the SEM data we collected. The algorithm demonstrated high fidelity, achieving accuracies of 0.99 on metal layers, 0.94 on doped layers, and 0.86 on polysilicon layers. While LASRE’s prioritization of connectivity over shape accuracy makes it exceptionally robust for HA applications, its high computational complexity makes it prohibitively slow for large-scale, continuous processing. Therefore, we utilize LASRE strictly as an offline automated annotator to generate a foundational set of image-mask pairs.
All experiments were conducted on an Intel Xeon Platinum 8570 CPU with 32GB of RAM and an NVIDIA B200 GPU with 180GB of GPU RAM.

\subsection{Learning the Mask Distribution with StyleGAN}
In the first phase of our methodology, the StyleGAN architecture was utilized to generate a dataset of synthetic binary masks. The model was trained to produce images at a final resolution of 256x256 pixels. The training followed the standard StyleGAN strategy of Progressive Growing, which begins at a low resolution (4x4 pixels) and incrementally adds layers to the generator and discriminator. This allows the model to learn large-scale features first before focusing on finer details, ensuring stable training. The process used an initial dataset of 505 real masks, and horizontal flipping was applied as data augmentation.

The training process was stabilized using the Wasserstein GAN with Gradient Penalty (WGAN-GP) loss function. The generator's ($G$) objective is to create images from a latent vector ($z$) that the critic (discriminator, $D$) scores as highly as real images. The generator's loss is defined as:
\begin{equation}
    L_{G} = - \mathbb{E}_{z}[D(G(z))]
\end{equation}
The critic's ($D$) objective is to maximize the difference between its scores for real images ($B$) and generated images ($G(z)$). Its loss function includes a crucial gradient penalty term to ensure stable training:
\begin{align}
L_{D} 
&= \mathbb{E}_{z}\!\left[D(G(z))\right] 
   - \mathbb{E}_{B}\!\left[D(B)\right] \notag \\
&\quad + \lambda_{GP}\, 
   \mathbb{E}_{\hat{B}}\!\left[\big(\lVert \nabla_{\hat{B}} D(\hat{B}) \rVert_2 - 1\big)^2\right]
\end{align}
The term $\hat{B}$ in the gradient penalty represents a sample created by interpolating between a real image ($B$) and a generated image ($G(z)$). It is calculated by choosing a random number $\epsilon$ from a uniform distribution between 0 and 1, and then applying the formula $\hat{B} = \epsilon \cdot B + (1 - \epsilon) \cdot G(z)$. By evaluating the critic's gradient at these intermediate points, the penalty enforces a soft constraint that is critical for stabilizing the training process.

To evaluate the impact of dataset size on model performance, we trained a series of models on datasets containing 10, 50, 100, 200, 300, 400, and 500 images. For this experiment, all training conditions and hyperparameters were held constant across every run to isolate the effect of dataset size. Each model was trained using the Adam optimizer with a learning rate of 0.0001, Adam betas of (0.0, 0.999), and a gradient penalty weight, $\lambda_{GP}$, of 10.0. The training schedule was also identical for all models, following a series of batch sizes (32, 32, 32, 16, 16, 8, 4) and epochs (60, 100, 100, 120, 120, 160, 300) for each resolution step. These settings follow established conventions for stable adversarial training rather than task-specific tuning. The gradient-penalty weight $\lambda_{GP}=10$ and the Adam configuration (learning rate $1\times10^{-4}$, $\beta_1=0$) are the standard values for gradient-penalty Wasserstein training~\cite{gulrajani2017improved}, where a reduced first-moment coefficient mitigates the instability that momentum can introduce into critic updates.
After training, the final generator was used to produce 5,000 synthetic binary masks for the project's second phase.

\subsection{Conditional Image Synthesis with Pix2PixHD}
In the second phase of our methodology, Pix2PixHD was trained using a dataset of 505 pairs of real SEM images and binary masks extracted from smart card ICs. The training objective was to teach the generator $G$ to create SEM images from input masks that are indistinguishable from real ones. To achieve this, two key loss functions from the Pix2PixHD framework were applied: an Adversarial Loss to ensure the realism of the output, and a Feature Matching Loss to improve the output's details by comparing intermediate feature maps from the discriminator. The model was trained using a dataset of 505 pairs of real SEM images and binary masks extracted from smart card ICs. The training process is adversarial, involving a generator ($G$) and discriminators ($D_k$). The $k$ number of discriminators is trained to maximize their ability to distinguish real image pairs from fake ones, following the objective function:
\begin{align}
L_{\text{GAN}}(G, D_k) 
&= \mathbb{E}_{(A,B)}\!\left[\log D(A, B)\right] \notag \\
&\quad + \mathbb{E}_{A}\!\left[\log \big(1 - D(A, G(A))\big)\right]
\end{align}
To stabilize training and enhance detail, a Feature Matching Loss, $L_{FM}(G, D_k)$, was used, which compares the intermediate feature maps from the multi-scale discriminator ($D_k$). It calculates the $L1$ distance between real pairs (mask $A$, SEM $B$) and fake pairs (mask $A$, generated SEM image $G(A)$):
\begin{align}
L_{\text{FM}}(G, D_k) 
&= \mathbb{E}_{(A,B)} 
   \sum_{i=1}^{T} \frac{1}{N_i} \notag \\
&\quad \times 
   \big\lVert D_k^{(i)}(A, B) - D_k^{(i)}(A, G(A)) \big\rVert_1
\end{align}
The generator's final objective minimizes a combination of the adversarial loss and this feature matching loss, with the weight for the feature matching loss, $\lambda_{FM}$, set to 10.0. To evaluate the impact of dataset size, we trained a series of pix2pixHD models on datasets of varying sizes: 10, 50, 100, 200, 300, 400, and 500 images. The model was trained for a total of 400 epochs using the Adam optimizer. The learning rate was set to 0.0002 with beta values of (0.5, 0.999), and a linear decay was applied after the first half of training. All training was conducted with a batch size of 32 using images at a 256x256 pixel resolution. The learning rate of $2\times10^{-4}$ with $\beta_1=0.5$ and linear decay over the second half of training, together with the feature-matching weight $\lambda_{FM}=10$, follow the reference Pix2PixHD configuration~\cite{wang2018high}, which we adopt to remain consistent with the regime under which the architecture was originally validated.
We evaluated the performance of the trained pix2pixHD models using Peak Signal-to-Noise Ratio (PSNR), Structural Similarity Index (SSIM) and Fréchet Inception Distance (FID). Please note that all evaluations in this phase were conducted by translating masks from our held-out validation set, as no corresponding ground truth images exist for the synthetically generated masks produced in Phase 1.

After training was complete, the optimized generator model was used to generate 5,000 corresponding synthetic SEM images from the 5,000 synthetic binary masks created in the previous step. Ultimately, this resulted in a complete synthetic dataset of 5,000 pairs that retains the characteristics of the real data while being free from privacy and intellectual property concerns.

\subsection{Training the Segmentation Network}
We employed the U-Net architecture, a convolutional network designed for precise biomedical image segmentation, to perform this task. The model consists of a symmetric encoder-decoder structure:
\begin{itemize}
    \item Encoder (Downsampling Path): The encoder is composed of 8 downsampling blocks (\texttt{UNetDown}) that use strided convolutions to progressively reduce spatial dimensions while extracting increasingly complex contextual features from the input SEM image.
    \item Decoder (Upsampling Path): The decoder path symmetrically uses transposed convolutions (\texttt{UNetUp}) to upsample the feature maps, gradually reconstructing a full-resolution output mask.
    \item Skip Connections: A key feature of the U-Net is the use of skip connections, which concatenate the feature maps from the encoder directly to the corresponding layers in the decoder. This allows the network to combine high-level semantic information from the decoder with fine-grained spatial details from the encoder, enabling highly accurate localization.
    \item Output Layer: The final layer uses a Sigmoid activation function to produce a pixel-wise probability map, where each pixel value represents the probability of belonging to the foreground mask.
\end{itemize}

\begin{figure*}[ht]
    \centering
    \begin{subfigure}[t]{0.3\textwidth}
        \centering
        \includegraphics[width=2in, height=2in, trim={0 0 0 0},clip]{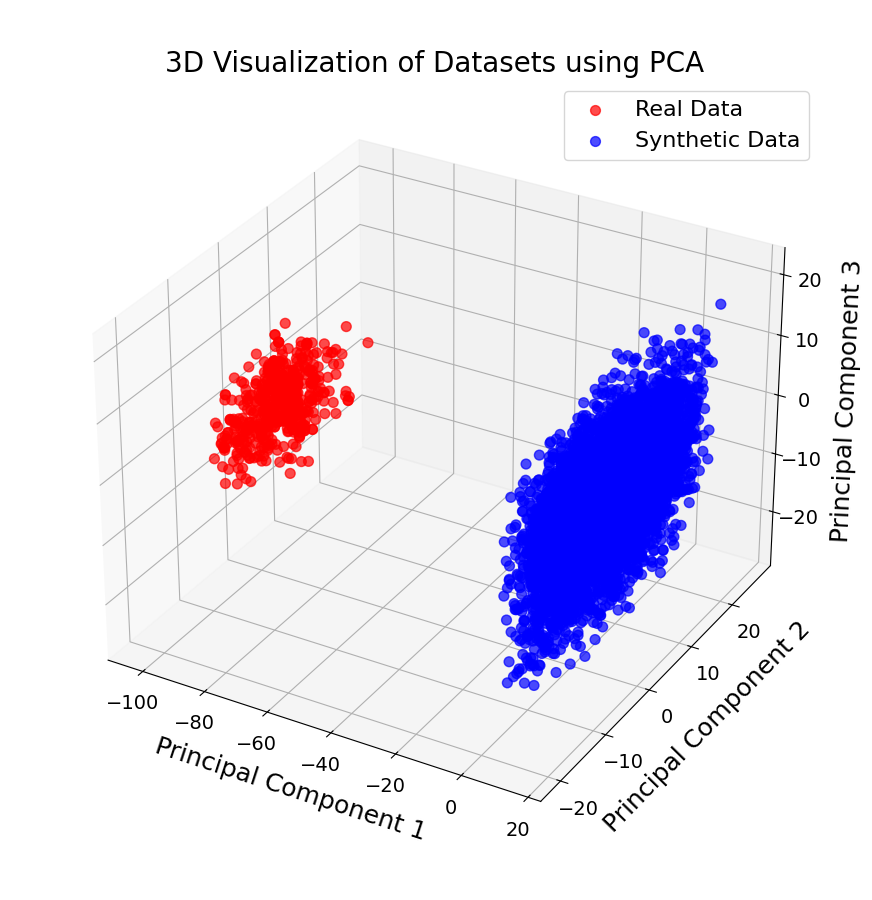}
        \caption{50 Images}
        \label{fig:pca_50}
    \end{subfigure}%
    ~ 
    \begin{subfigure}[t]{0.3\textwidth}
        \centering
        \includegraphics[width=2in, height=2in, trim={0 0 0 0},clip]{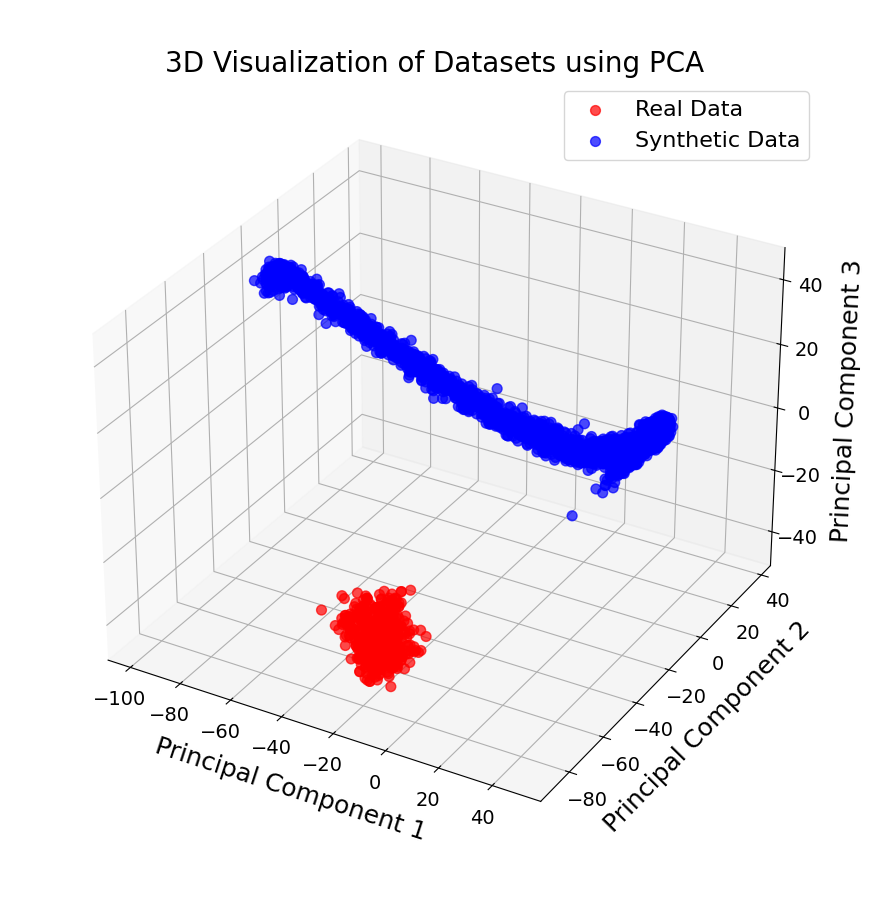}
        \caption{300 Images}
        \label{fig:pca_300}
    \end{subfigure}
    \begin{subfigure}[t]{0.3\textwidth}
        \centering
        \includegraphics[width=2in, height=2in, trim={0 0 0 0},clip]{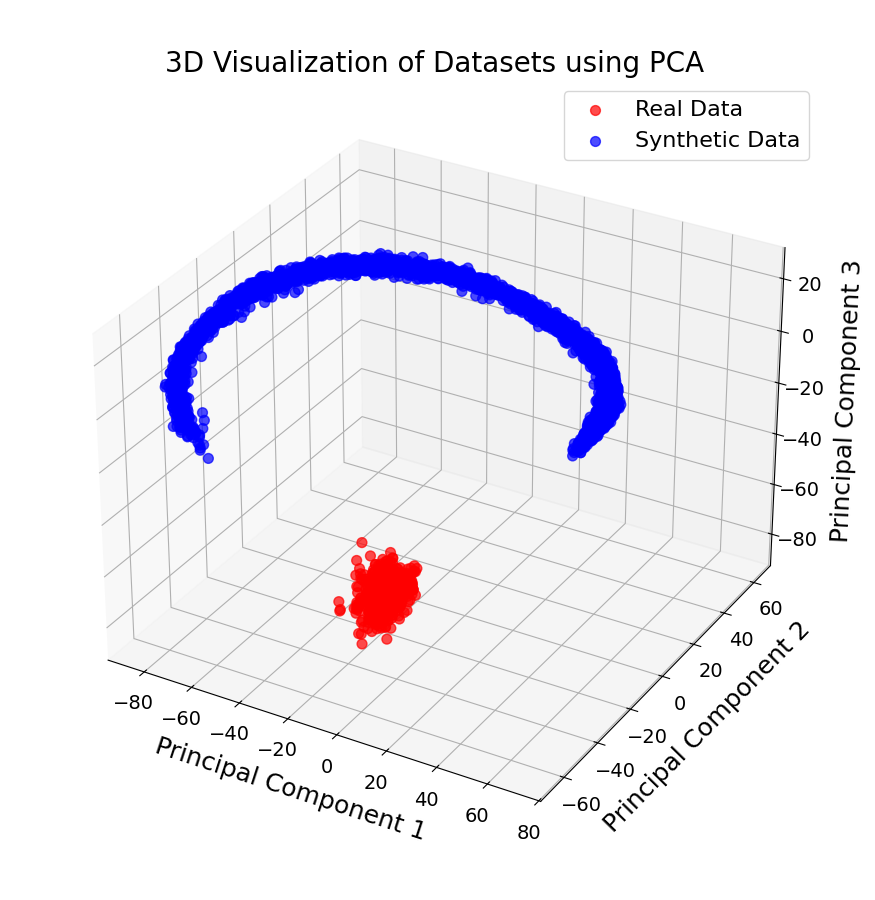}
        \caption{500 Images}
        \label{fig:pca_500}
    \end{subfigure}%
    
    \caption{3D PCA visualization for the metal layer comparing the real data distribution (red) with the synthetic data distribution (blue) from models trained on different dataset sizes. (a) The 50-image model produces a simple, dense cluster, indicating limited diversity. (b) The 300-image model marks a turning point, with the synthetic data forming a complex manifold. (c) The 500-image model refines this structure, showing the highest diversity.}
    \label{fig:pca_results}
\end{figure*}
The U-Net model was trained from scratch using the 5,000 synthetic image-mask pairs generated in the previous stage. The key training hyperparameters were as follows:
\begin{itemize}
    \item \textbf{Loss Function:} The model was optimized using Binary Cross-Entropy with Logits Loss (\texttt{BCEWithLogitsLoss}), which is well-suited for binary segmentation tasks.
    \item \textbf{Optimizer:} We used the Adam optimizer with a learning rate of 0.0002 and betas of (0.5, 0.999).
    \item \textbf{Training Parameters:} The model was trained for 20 epochs with a batch size of 64. All images were resized to a uniform 256x256 pixels for training.
\end{itemize}
We retain the same Adam configuration (learning rate $2\times10^{-4}$, $\beta_1=0.5$) used for the image-translation stage, so that the optimizer regime is consistent across the synthesis and segmentation networks.
After training on the synthetic data, the model's generalization capability was assessed on a test set composed of real SEM images and their ground truth masks. During evaluation, the model's Sigmoid output, which is a probability map, was converted into a binary mask by applying a threshold of 0.5 to each pixel. To quantitatively measure the model's performance, we used two standard segmentation metrics: the Dice Similarity Coefficient (Dice Score) and Intersection over Union (IoU). These metrics evaluate the overlap between the predicted mask and the ground truth mask, providing a robust measure of segmentation accuracy. For comparison, we also trained an identical U-Net architecture on a dataset of 350 real dataset, using the same training hyperparameters. This model serves as our performance baseline.

\subsection{Multi-faceted Privacy Preservation Analysis}
\label{sec/privacy_methods}
To quantitatively validate the privacy-preserving capabilities of our synthetic generation process, we performed a three-pronged analysis. Each method tests a different potential privacy failure, from broad distributional similarity to specific instance leakage. All analyses were conducted on 256x256 binary images.
For our privacy analysis, we define a threat model: the generative models and the original SEM data remain entirely private, while only the downstream segmentation model is considered public. 
In this context, our definition of ``privacy-preserving" centers on distorting the proprietary design information while preserving the visual and feature-based similarities required to train the segmentation network. As demonstrated in our quantitative results, Principal Component Analysis (PCA) reveals a clear spatial displacement between the real and synthetic datasets, forming distinct, non-overlapping clusters. This empirical divergence confirms that the synthetic distribution is novel in pixel space and does not replicate the original feature space. The absence of the proprietary routing context follows instead from the generation process itself: because the StyleGAN stage resamples the macroscopic layout rather than reproducing any specific real arrangement, the global routing that defines the original design is not preserved. The PCA separation is therefore evidence of distributional novelty which, together with this design-distortion argument, supports our privacy claim. Consequently, any privacy attacks—such as membership inference or gradient inversion—can only be executed against the public segmentation model. Because the underlying synthetic training data lacks actionable electrical intelligence and does not reconstruct original samples, even a successful data extraction yields layouts that are meaningless for reverse engineering the true design. This guarantees that the segmentation model can be safely shared without the concern of leaking the original intellectual property.

\vspace{0.2em}
\noindent\textbf{Distributional Privacy via PCA}
The first test assesses privacy at a high level by checking if the synthetic and real data distributions are trivially identical. Each image was flattened into a 65,536-dimensional vector and reduced to its first three principal components using PCA. A clear visual separation in the resulting 3D scatter plot shows that the synthetic data does not simply replicate the original data's feature space, providing a foundational check against simple reconstruction.

\vspace{0.2em}
\noindent\textbf{Defense Against MIA via Classification}
The second test provides evidence for the method's resilience to Membership Inference Attacks (MIA). An MIA's goal is to determine if a specific sample was present in a model's training set. Our approach defends against this by replacing the original private dataset ($R$) with a synthetic one ($S$) for model training. Consequently, an MIA on a downstream model can, at best, only infer membership in the non-private synthetic set $S$, revealing no information about the contents of $R$.
To validate this defense, we conducted a classification test to prove that S is statistically distinct from R. After correcting for dataset imbalance by subsampling, a Logistic Regression classifier was trained on a balanced 1:1 dataset. High classification accuracy (approaching 100\%) confirms that the two domains are easily separable, thereby validating that membership in $S$ is not a proxy for membership in $R$ and the core premise of the MIA is successfully thwarted.

\vspace{0.2em}
\noindent\textbf{Instance-Level Privacy Analysis}
The final and most direct privacy test searches for data leakage and 1-to-1 memorization. For each real training image, we searched the entire synthetic dataset to find its closest counterpart using Hamming distance (HD). The goal was to determine if any synthetic image is a near-perfect copy of a real one. The similarity of these closest pairs was then quantified using the Dice Similarity Coefficient and Intersection over Union (IoU). Low similarity scores across the dataset provide strong evidence that the synthetic images are novel creations, preserving instance-level privacy by not reconstructing original training samples.

\section{Results}
\label{sec:results}

\subsection{Impact of Dataset Size on Generation Quality}
Our investigation began by evaluating the effect of training dataset size on the StyleGAN model's ability to generate synthetic masks. We trained separate models on datasets ranging from 10 to 500 images and observed a strong correlation between data quantity and generator performance.

The model trained on a minimal dataset of 10 images failed to converge, producing unusable, low-fidelity outputs. Increasing the dataset to 50 images allowed the model to learn and produce coherent mask structures, but with limited diversity. PCA was performed to visually analyze this progression in Figure \ref{fig:pca_results}.

For clarity, we have selected the figures that represent the most significant milestones. The plots for the 100, 200, and 400-image models were omitted, as they are intermediate steps that show gradual improvement but not a fundamental change in the data's structure.

\begin{itemize}
    \item Figure \ref{fig:pca_50} confirms the Partial Success of the 50-image model, where the generated samples are confined to a dense, simple cluster, indicating limited diversity.
    \item Figure \ref{fig:pca_300} shows a Turning Point in Diversity with the 300-image model. The distribution of synthetic data transforms from a simple cluster into a distinct, curved manifold, marking the point where the generator captures a more complex structure.
    \item Figure \ref{fig:pca_500} represents the Highest Diversity Achieved with the 500-image model. This structural complexity is further refined, spanning a wider and more defined arc in the feature space.
\end{itemize}

\subsection{Qualitative Analysis of Generation Quality}
\begin{figure}[t]
\centering
\captionsetup{justification=centering}
\includegraphics[width=1\linewidth]{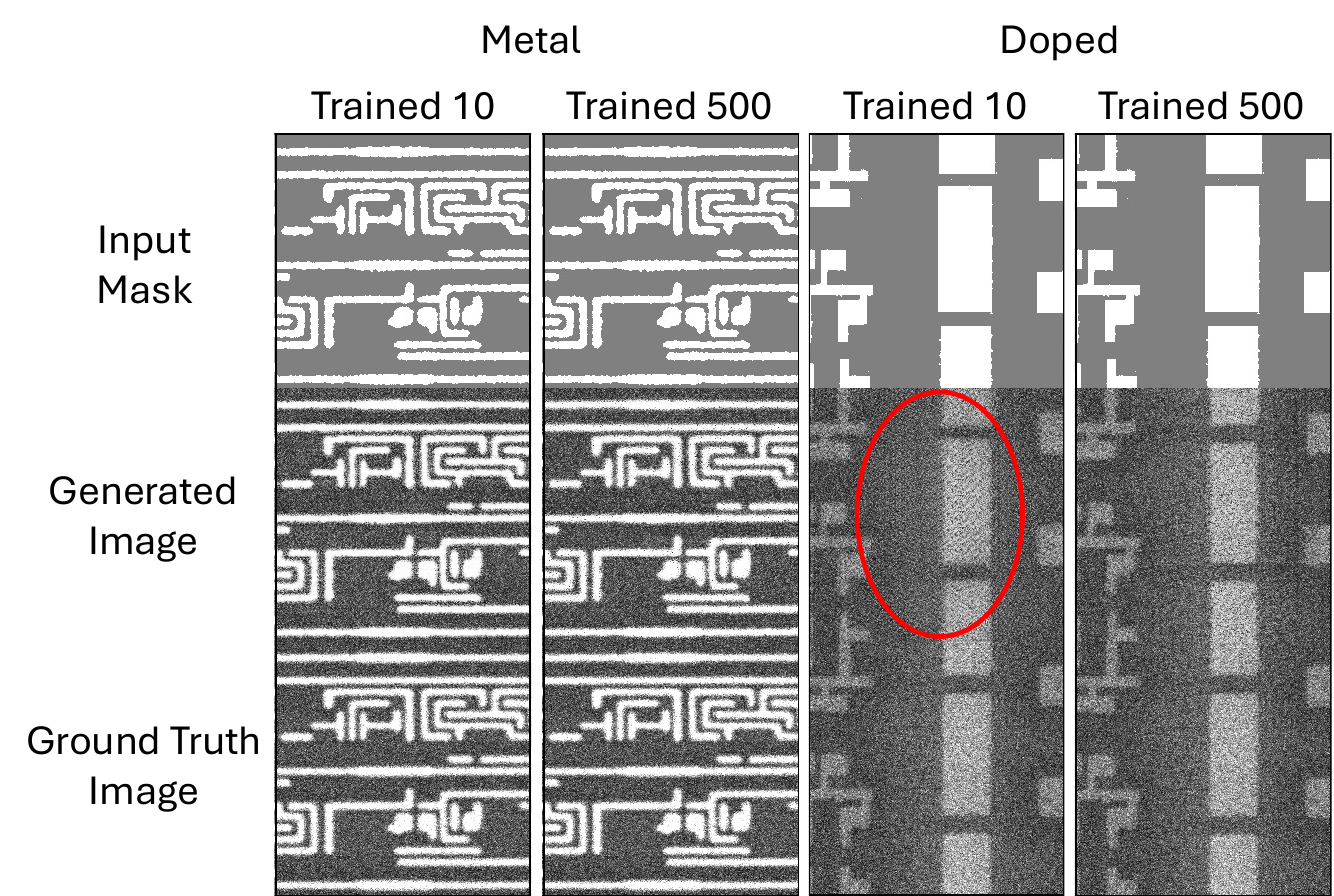}
\caption{Synthetic metal and doped SEM images.}
\label{fig:pix2pixhd}
\end{figure}

\begin{table}[t]
\centering
\caption{Generation quality comparison by layout and dataset size.}
\label{tab:pix2pix_results}
\resizebox{\columnwidth}{!}{
\begin{tabular}{lcccc}
\toprule
\textbf{Layout} & \textbf{Dataset Size} & \textbf{Avg. PSNR $\uparrow$} & \textbf{Avg. SSIM $\uparrow$} & \textbf{Avg. FID $\downarrow$} \\
\midrule
\multirow{2}{*}{Metal} & 10  & 15.2068 & 0.3826 & 46.7116\\
                       & 500 & 15.2018 & 0.3775 & 43.6454\\
\midrule
\multirow{2}{*}{Doped} & 10  & 14.2506 & 0.0544 & 60.014\\
                           & 500 & 14.1932 & 0.0560 & 43.7201\\
\midrule
\multirow{2}{*}{Polysilicon} & 10  & 15.5244 & 0.1579 & 62.2876\\
                           & 500 & 16.3462 & 0.2114 & 35.7038\\
\bottomrule
\end{tabular}
}
\end{table}
To complement the quantitative metrics, we performed a qualitative visual analysis of the images generated by the pix2pixHD models. We used samples from our validation set to assess the models' ability to translate both simple (metal layers) and complex (doped layers) structures.

\begin{figure}[t]
\centering
\captionsetup{justification=centering}
\includegraphics[width=1\linewidth]{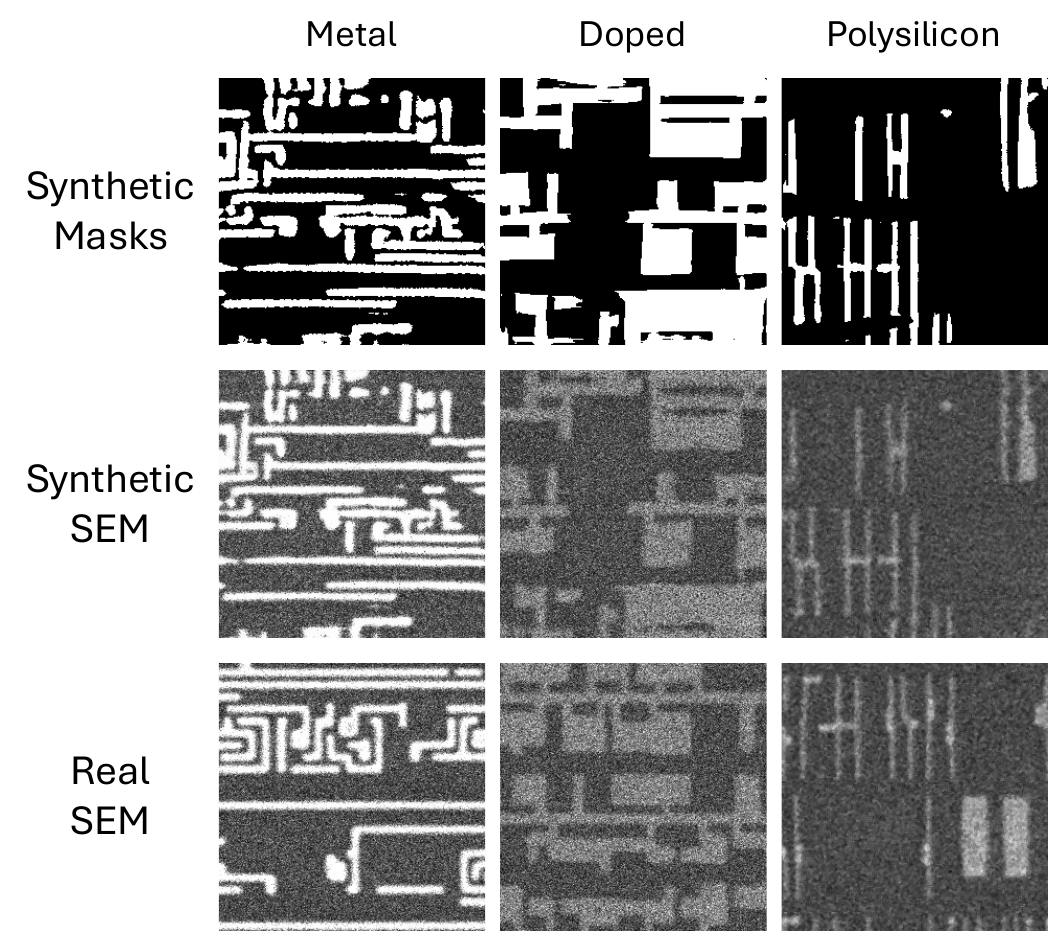}
\caption{Samples of synthetic datasets.}
\label{fig:synthetic}
\vspace{-9.5mm}
\end{figure}

Our findings show that for relatively simple structures, such as metal layers, the model trained on only 10 images produces remarkably good results.
As seen in Figure~\ref{fig:pix2pixhd}, the generated output is visually coherent and largely accurate compared to the ground truth. This confirms that the fundamental image-to-image mapping can be learned from a very small dataset. However, for more intricate patterns such as doped layers, the benefits of a larger dataset become clear. While both results are plausible, the model trained on 500 images demonstrates superior performance in rendering fine details and maintaining edge fidelity, as highlighted by the red circles. This indicates that while a small dataset is sufficient for learning the general structure, a larger dataset is crucial for the model to learn and accurately reproduce fine-grained details. 
The quantitative performance of the models was evaluated on our held-out validation set using PSNR, SSIM, and FID, with the results summarized in Table~\ref{tab:pix2pix_results}. The absolute SSIM scores remain predictably low due to the inherent visual characteristics of SEM imagery. Unlike standard datasets with uniform backgrounds, real and synthetic SEM images are characterized by persistent, high-frequency ``salt-and-pepper" background noise. Because pixel-based metrics like SSIM heavily penalize pixel-wise misalignments in these uncorrelated noise patterns—as well as the fuzzy boundaries typical of doped layers—they do not accurately reflect the generative quality. Furthermore, these metrics show only a marginal difference between models trained on 10 images versus 500 images, suggesting that the rigid, macroscopic spatial structures are learned very early during training. However, the inclusion of the FID score reveals a significant improvement in perceptual quality and feature distribution when scaling the dataset. This divergence is most pronounced in the polysilicon and doped layer layouts; for instance, the polysilicon model's FID score drops dramatically from 62.2876 to 35.7038. This demonstrates that while small datasets can capture basic structural alignments, larger datasets are crucial for refining complex, high-frequency textures—such as the characteristic SEM noise—and achieving a highly realistic generative distribution that traditional pixel-wise metrics fail to evaluate properly. As illustrated by the samples in Figure~\ref{fig:synthetic}, the generated synthetic images of all different layers exhibit a high degree of visual fidelity and are very similar to the real SEM data, validating our generation process.

\subsection{Results of the Segmentation Task}
The primary objective of this phase was to assess if the U-Net model, trained exclusively on the 5,000 synthetic image-mask pairs, could generalize to segment real SEM images. The trained model was therefore evaluated on a held-out test set composed of real SEM images and their corresponding ground truth masks.

\begin{figure}[ht]
\centering
\captionsetup{justification=centering}
\includegraphics[width=0.99\linewidth]{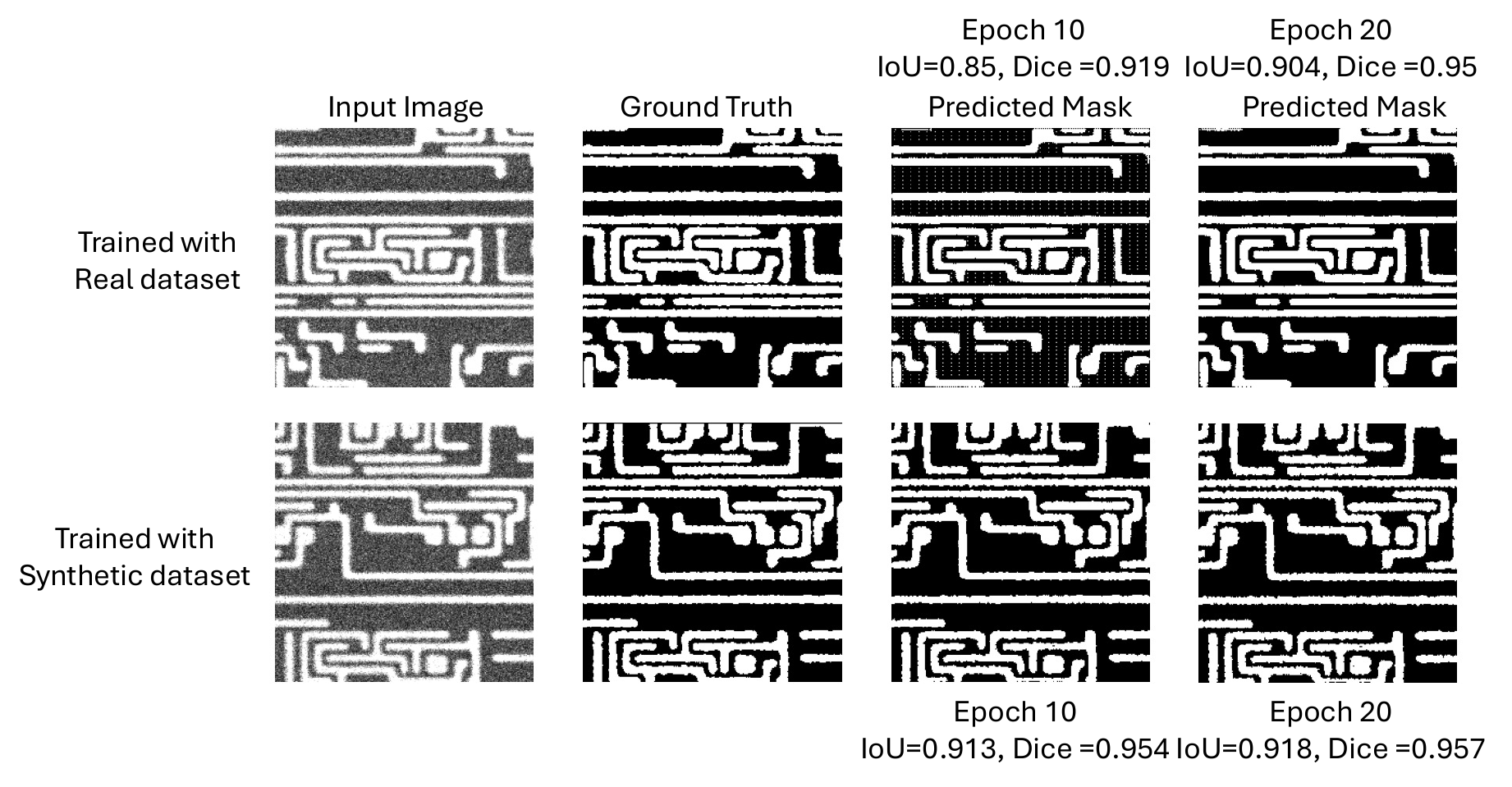}
\caption{Segmentation Performance}
\label{fig:segmentation_results}
\end{figure}

The model trained on synthetic data demonstrated excellent generalization capability, accurately segmenting real images despite having never been trained on them. As shown in Figure~\ref{fig:segmentation_results}, a qualitative and quantitative analysis reveals that the model's performance was highly successful, even showing advantages over a baseline model trained on a real dataset.

Our model trained exclusively on synthetic data yielded highly successful results, achieving a final Intersection over Union (IoU) of 0.918 and a Dice Score of 0.957. This performance not only surpassed our own baseline model trained on real data (IoU 0.904, Dice 0.95) but is also highly competitive with state-of-the-art benchmarks. For comparison, \cite{wilson2021refics} reports a top IoU of 0.95 for the CBDNet model, which was trained on a large dataset of 90,000 synthetic images. Achieving near-benchmark performance with a much smaller, entirely synthetic dataset underscores the power and practicality of our privacy-preserving approach.

Furthermore, a visual comparison at an earlier training stage highlights a more significant difference. After only 10 epochs, the prediction from the model trained with the real dataset exhibited noticeable artifacts, such as dots in the predicted mask, and achieved a lower IoU of 0.85. In contrast, the model trained with the synthetic dataset produced a cleaner prediction at 10 epochs and had already reached a higher IoU of 0.913. This suggests that training on the larger synthetic dataset may lead to more stable and efficient convergence. 

This finding is significant for several key reasons. First, it accomplishes two important technical goals:
\begin{itemize}
\item It validates the quality and realism of the synthetic data generated by the StyleGAN and pix2pixHD pipeline in the previous phases.
\item It demonstrates a successful ``sim-to-real" transfer, proving that a segmentation network can be effectively trained on synthetic data for this task.
\end{itemize}

Most importantly, this approach provides a robust, privacy-preserving workflow. By training exclusively on synthetic data, it avoids the need for laborious manual annotation of potentially sensitive real images. Furthermore, this method inherently defends against data leakage from sophisticated privacy attacks. For instance, attacks like membership inference, which aims to determine if a specific person's data was in the training set, or gradient inversion, which can reconstruct training images from model updates, would fail to expose real information. Since the model is trained exclusively on synthetic data, these attacks could, at best, only reveal the synthetic images, ensuring the privacy of the original data source is preserved.

\subsection{Privacy}
Following the methodology outlined in Section \ref{sec/privacy_methods}, we performed a multi-faceted analysis to evaluate the privacy guarantees and distributional characteristics of the synthetic dataset compared to the original real images.

\begin{figure}[t]
\centering
\captionsetup{justification=centering}
\includegraphics[width=0.6\linewidth,trim={2cm 4cm 0 3cm},clip]{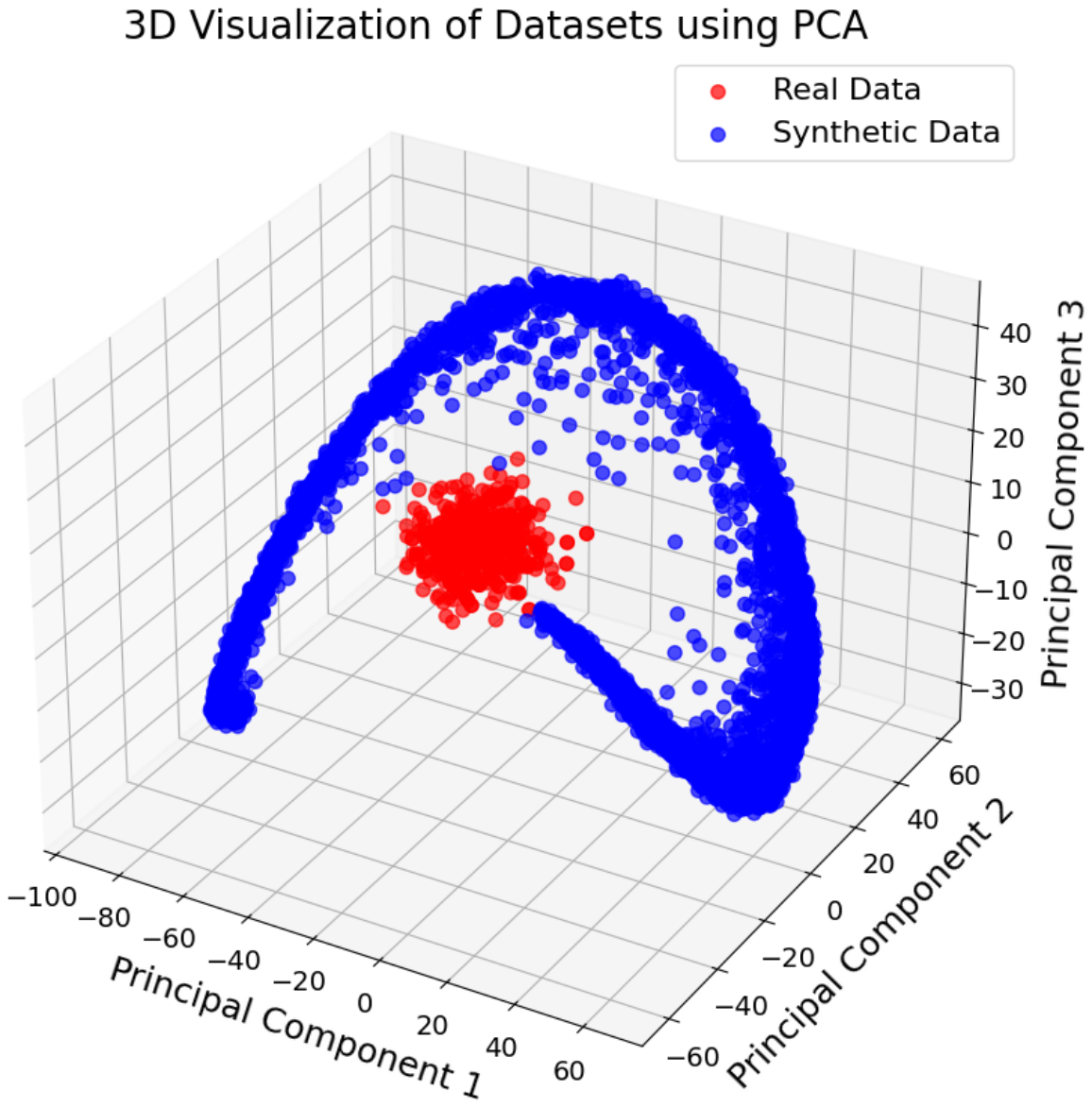}
\caption{Real vs. synthetic metal layer SEM image distributions.}
\label{fig:pca}
\end{figure}


\begin{table*}[h!]
\centering
\caption{Dice and IoU metrics for different material layers.}
\label{tab:dice_iou_metrics}
\begin{tabular}{l ccc ccc}
\toprule
& \multicolumn{3}{c}{\textbf{Dice Coefficient}} & \multicolumn{3}{c}{\textbf{IoU}} \\
\cmidrule(lr){2-4} \cmidrule(lr){5-7}
           & \textbf{Min} & \textbf{Mean} & \textbf{Max} & \textbf{Min} & \textbf{Mean} & \textbf{Max} \\
\midrule
Metal      & 0.3904 & 0.4877 $\pm$ 0.0353 & 0.5963 & 0.2425 & 0.3232 $\pm$ 0.0309 & 0.4248 \\
Doped  & 0.1969 & 0.2586 $\pm$ 0.0319 & 0.3981 & 0.1092 & 0.1532 $\pm$ 0.0218 & 0.2485 \\
Polysilicon& 0.0114 & 0.0725 $\pm$ 0.0406 & 0.3308 & 0.0057 & 0.0381 $\pm$ 0.0230 & 0.1982 \\
\bottomrule
\end{tabular}
\vspace{-3mm}
\end{table*}

\begin{itemize}
    \item Distributional Privacy via PCA: The Principal Component Analysis, conducted on the full datasets (505 real and 5,000 synthetic images), revealed a clear and significant separation between the two data distributions. As illustrated in Figure \ref{fig:pca}, the real and synthetic data form two distinct, non-overlapping clusters in the 3D space defined by the first three principal components. The cluster representing the synthetic data is visibly denser, reflecting its larger size. This spatial displacement provides strong visual evidence that the synthetic generation process produced a novel data distribution that does not simply replicate the original feature space, serving as a foundational confirmation of distributional privacy.

    \item Defense Against MIA via Classification: To formally measure the statistical separability, a Logistic Regression classifier was trained on a balanced dataset of 505 real and 505 randomly subsampled synthetic images. The model demonstrated extremely high performance in distinguishing between the two classes on an unseen test set, achieving a final accuracy of 100\% with precision and recall scores of 1.00 and 1.00 for the real and synthetic class, respectively. This perfect ability to separate the two datasets provides strong evidence of their statistical divergence, which validates that membership in the synthetic set is not a proxy for membership in the original private set. This effectively thwarts the goal of a membership inference attack, as any successful inference would only reveal membership in the non-private synthetic data, not the original private samples.
    
    \item Instance-Level Privacy: To test for direct data leakage or memorization, a nearest neighbor analysis was performed to find the most similar synthetic image for each of the 505 real images. Table \ref{tab:dice_iou_metrics} shows comparisons of the metal layer that yielded a mean Dice Similarity Coefficient of 0.4877 (SD = 0.0353) and a mean Intersection over Union (IoU) of 0.3232 (SD = 0.0309).
    Most critically for privacy, the maximum observed similarity between any real-synthetic pair was 0.5963 for the Dice score and 0.4248 for the IoU score. The absence of any similarity score approaching 1.0 confirms that no synthetic image is a direct copy or near-perfect reconstruction of a real training sample. This provides strong evidence for instance-level privacy and the novelty of the generated images.
\end{itemize}

Our results demonstrate that the generated dataset is statistically distinct from and not a direct reconstruction of the original data. A significant implication of this finding is the enhanced privacy of downstream models trained on this synthetic data. Specifically, our methodology provides a strong defense against privacy breaches from gradient inversion attacks. Even if an attacker could successfully mount such an attack on a model trained with our data, the reconstructed image would, at best, be a sample from the synthetic distribution, not the original, private data. Because we have already proven the synthetic images are novel and not copies, this attack vector is effectively neutralized. This ensures our approach provides an end-to-end privacy pipeline, protecting not only the original dataset but also the models subsequently trained on its synthetic version.

Furthermore, while image-level metrics confirm global novelty, we must consider the worst-case scenario regarding local memorization: generative models may occasionally produce small localized patches that perfectly resemble real data. This is expected, as the model must learn standard, non-proprietary building blocks, such as basic transistors in doped layers or minimum-width routing. However, even in this worst-case scenario where a small, localized patch of the synthetic data matches the real dataset, this isolated geometry provides zero functional intelligence without the broader routing context. Because our macro-level synthetic layouts are demonstrably novel, the risk of exposing actionable, proprietary circuit functions through these minor localized similarities is effectively eliminated.

A distinct question, separate from the memorization analysis above, is whether the generator might coincidentally produce a synthetic layout that is itself electrically coherent, even though it copies no real sample. We emphasize that electrical coherence alone does not constitute an IP breach: the proprietary information in an IC resides in one \emph{specific} circuit, so for a synthetic layout to expose that IP it would have to reproduce that particular design. Our instance-level analysis shows that no synthetic layout is a near-copy of a real sample (the maximum observed similarity across the dataset is 0.5963 Dice), and our distributional analyses show that the synthetic and real sets are statistically separable rather than replicas of one another. Consequently, even a coincidentally coherent synthetic layout would represent a \emph{different} circuit, not the protected one, and would not leak the original design. This is consistent with, but stronger than, the local-memorization argument above, in which any small repeated structure corresponds only to standard, non-proprietary building blocks. We acknowledge, however, that these analyses quantify novelty in image space rather than electrical validity directly. For workflows that publish the synthetic layouts themselves---most notably the public benchmark repository proposed in Section~\ref{sec:futureworks}---an explicit validation stage would further strengthen the guarantee, for instance by applying design-rule or connectivity checks (e.g., netlist extraction) to flag any layout that inadvertently forms a coherent circuit, or by training a classifier for the same purpose. A systematic study of how often such coincidental coherence arises is left to future work.
\section{Future Works}
\label{sec:futureworks}
Future work can extend this research along three primary avenues: enhancing the quality of the synthetic data, broadening its applicability across more complex and varied imagery, and strengthening the privacy guarantees. First, future work could explore advanced generative models to create a dataset that is both highly realistic and more diverse than the limited source data. Instead of aiming to perfectly replicate the original distribution, the goal would be to synthesize novel yet plausible examples, thereby training a more robust downstream segmentation model and potentially boosting its generalization performance on unseen cases. This involves investigating the relationship between the information density of source images and the complexity of the PCA data manifold to predict the minimum required dataset size. A critical next step is to analyze if this increased diversity improves the functional correctness of the final segmentation by quantifying whether the varied training data reduces critical errors, such as short-circuits.

A current limitation of our evaluation is that it is restricted to a single smart-card IC imaged under one acquisition configuration, where the layers of interest are relatively simple and exhibit strong contrast against their surroundings. The unsupervised annotation step reflects this dependence: LASRE accuracy declines from 0.99 on metal layers to 0.94 on doped and 0.86 on polysilicon, indicating that intensity-based grouping degrades as geometric complexity increases and feature--background contrast weakens. Extending the pipeline to more complex and varied imagery---across additional technology nodes, manufacturers, and device types---is therefore an important direction. One promising avenue is to treat acquisition variability as a domain-randomization signal within the Pix2PixHD stage, deliberately synthesizing the texture and noise characteristics of different imaging conditions so that the downstream segmentation model becomes robust to them rather than tuned to a single setup.

Beyond model design, the acquisition settings of the source SEM images directly influence downstream performance, since the entire pipeline is conditioned on these initial real images. For the smart card IC studied here, images were acquired with a secondary-electron (SE) detector at an accelerating voltage of 5\,keV, a field of view of 200\,$\mu$m, and a dwell time of 10\,$\mu$s per pixel (Section~\ref{sec:methodology}), a configuration that was sufficient to support the segmentation performance reported in our experiments (IoU 0.918, Dice 0.957). We deliberately avoid prescribing a single universal recipe, as the optimal detector, accelerating voltage, and brightness/contrast settings vary substantially from one device and layer stack to another. Instead, the operative requirement is that the source images be clear: the features of interest should be well distinguished from the background and free of excessive noise, with brightness and contrast set to span the available dynamic range without clipping. Provided this clarity is met, our pipeline can synthesize high-quality images and train an effective segmentation model regardless of the specific instrument configuration used to obtain the seed data. A systematic characterization of how degraded or noisier source acquisitions affect the synthetic data and the resulting segmentation accuracy is a useful direction for future study.

To broaden the impact of this work, we plan to leverage our pipeline to create and host a large-scale public repository of synthetic SEM images, allowing the community to benchmark AI algorithms for hardware assurance without requiring access to proprietary IP. Finally, to achieve even higher levels of privacy, our synthetic data approach could be combined with other Privacy-Preserving techniques (PPTs). For instance, applying PPTs during the generator's training could provide stronger mathematical guarantees against data leakage from the original sensitive data. Furthermore, using the synthetic data within a Federated Learning (FL) framework would enable collaborative training. This is a significant advantage, as it allows clients to participate in the FL process without the worry of common privacy attacks (like gradient inversion) exposing their sensitive local data. This secure foundation for collaboration would not only provide multiple layers of robust privacy protection but would also allow the segmentation model to become more generalized and powerful by learning from the diverse datasets contributed by all participants.
\section{Conclusion}
\label{sec:conclusion}
In this work, we addressed the critical challenge of training deep learning models for SEM image segmentation of ICs when access to real data is limited by strict intellectual property (IP) and confidentiality constraints. We validated a pipeline to generate 5,000 synthetic images and showed that a U-Net trained on this data outperformed a baseline model trained on a smaller set of 350 real images. The primary contribution of this research is a method that addresses three critical challenges in training deep learning models for hardware assurance on sensitive data. First, it mitigates IP exposure risks from attacks like gradient inversion and membership inference; in a practical deployment where only the trained model is shared, any successfully extracted synthetic data is statistically distinct from the original layouts and reproduces no specific proprietary routing, so it cannot be used to reverse-engineer the protected design. Second, it surpasses the performance of traditional training when real data is scarce. Third, it substantially reduces the burden of manual annotation. Ultimately, this work validates synthetic data generation as a powerful, secure, and practical solution for high-performance deep learning in this domain.

{
    \small
    \bibliographystyle{ieeenat_fullname}
    \bibliography{main}
}

\end{document}